\documentclass[twocolumn,letterpaper,accepted=2020-02-05]{quantumarticle}
\pdfoutput=1
\usepackage[utf8]{inputenc}
\usepackage[english]{babel}
\usepackage[T1]{fontenc}
\usepackage{amsmath}
\usepackage{hyperref}
\usepackage{tikz}
\usepackage{lipsum}
\usepackage[numbers,sort&compress]{natbib}

 \usepackage{verbatim}
 \usepackage{amssymb}
 \usepackage{amsthm}
 \usepackage{latexsym}
 \usepackage{amsfonts}
 \usepackage{epsfig}
 \usepackage{epstopdf}
 \usepackage{wasysym} % for certain special symbols
 \usepackage{color}
 \definecolor{darkblue}{rgb}{0,0,.5}
 \usepackage[all]{hypcap}
 \usepackage[makeroom]{cancel}

\newcommand{\C}[1]{{\cal{#1}}}
\newcommand{\bb}[1]{\textbf{#1}}

\newcommand{\mf}[1]{{\mathfrak{#1}}}

\newcommand{\lr}[1]{{\left\langle {#1}\right\rangle}}
\newcommand{\rl}[0]{{\rangle\langle}}

\begin{document}

\title{Thermodynamics of Quantum Causal Models: An Inclusive, Hamiltonian Approach}

\author{Philipp Strasberg}
\affiliation{F\'isica Te\`orica: Informaci\'o i Fen\`omens Qu\`antics, Departament de F\'isica, Universitat Aut\`onoma de Barcelona, 08193 Bellaterra (Barcelona), Spain}
\orcid{0000-0001-5053-2214}

%\date{\today}

\begin{abstract}
 Operational quantum stochastic thermodynamics is a recently proposed theory to study the thermodynamics of open 
 systems based on the rigorous notion of a quantum stochastic process or quantum causal model. In there, a stochastic 
 trajectory is defined solely in terms of experimentally accessible measurement results, which serve as the basis to 
 define the corresponding thermodynamic quantities. In contrast to this observer-dependent point of view, a `black box', 
 which evolves unitarily and can simulate a quantum causal model, is constructed here. The quantum 
 thermodynamics of this big isolated system can then be studied using widely accepted arguments from statistical 
 mechanics. It is shown that the resulting definitions of internal energy, heat, work, and entropy have a natural 
 extension to the trajectory level. The canonical choice of them coincides with the proclaimed definitions of 
 operational quantum stochastic thermodynamics, thereby providing strong support in favour of that novel framework. 
 However, a few remaining ambiguities in the definition of stochastic work and heat are also discovered and in light of 
 these findings some other proposals are reconsidered. Finally, it is demonstrated that the first and second law hold 
 for an even wider range of scenarios than previously thought, covering a large class of 
 quantum causal models based solely on a single assumption about the initial system-bath state. 
\end{abstract}

%%%%%%%%%%%%%%%%%%%%%%%%%%%%%%%%%%%%%%%%%%%%%%%%%%%%%%%%%%%%%%%%%%%%%%%%%%%%%%%%%%%%%%%%%%%%%%%%%%%%%%%%%%%%%%%%%%%%%%%%
\section{Introduction}

The success of the classical framework of stochastic thermodynamics is undeniable. It pushes the validity of the laws 
of thermodynamics far beyond their original scope, it allows to consistently describe the thermodynamics of small 
fluctuating out-of-equilibrium systems, even along a single trajectory, and many of its predictions have been verified 
experimentally~\cite{BustamanteLiphardtRitortPhysTod2005, SekimotoBook2010, JarzynskiAnnuRevCondMat2011, SeifertRPP2012, 
VandenBroeckEspositoPhysA2015, CilibertoPRX2017}.

In contrast, how to describe the thermodynamics of small quantum systems along a single `trajectory' remains a subject 
of debate since 20 years. Obviously, the reason is the measurement backaction of an external observer, who manipulates 
a small quantum system and thereby \emph{changes} the process. This implies that any theory of quantum stochastic 
thermodynamics should be able to consistently treat the measurement backaction and is necessarily different from its 
classical counterpart~\cite{PerarnauLlobetEtAlPRL2017}. Over the past, many different approaches have been put forward, 
often differing in their predictions and lacking either an experimentally feasible way to verify them or the ability to 
describe quantum effects. Recently, based on a rigorous notion of a quantum stochastic process or quantum causal 
model~\cite{LindbladCMP1979, AccardiFrigerioLewis1982, ChiribellaDArianoPerinottiPRA2009, HardyRSocA2012, 
OreshkovCostaBruknerNC2012, CostaShrapnelNJP2016, OreshkovGiarmatziNJP2016, PollockEtAlPRA2018, MilzEtAlArXiv2017}, an 
`operational' approach to quantum stochastic thermodynamics was constructed~\cite{StrasbergPRE2019, 
StrasbergWinterPRE2019, StrasbergPRL2019}. It puts the experimenter in the foreground by explicitly including all 
external interventions (state preparation, measurements, feedback operations, etc.) in the description. A `stochastic 
trajectory' is defined solely in terms of experimentally available (classical) measurement results, on which the 
corresponding thermodynamic quantities are built. The formalism is free from many restrictive and previously used 
assumptions (e.g., perfect measurements, continous measurements, detailed control about the bath degrees of freedom, no 
feedback control, use of ambiguous notions for time-reversed trajectories, etc.) and can be readily applied to analyse 
a multitude of experiments including Refs.~\cite{SayrinEtAlNature2011, ZhouEtAlPRL2012}. 

Nevertheless, the definitions used in Refs.~\cite{StrasbergPRE2019, StrasbergWinterPRE2019, StrasbergPRL2019} were 
derived from an observer-dependent point of view, involving quantum measurement theory, subjective choices of the 
`Heisenberg cut', and certain classicality assumptions. To circumvent the use of any such elements, this paper rederives 
the framework of operational quantum stochastic thermodynamics based on an inclusive, Hamiltonian (`autonomous') 
approach. By using only arguments from nonequilibrium statistical mechanics of isolated systems, we provide a solid and 
independent justification for the definitions of Refs.~\cite{StrasbergPRE2019, StrasbergWinterPRE2019, StrasbergPRL2019}. 
To the best of the authors' knowledge, this distinguishes the operational approach from other proposals in quantum 
stochastic thermodynamics. 

The idea to model everything autonomously is not novel and has been used very successfully to understand the physics of 
Maxwell's demon~\cite{JacobsPRA2009, StrasbergEtAlPRL2013, HartichBaratoSeifertJSM2014, HorowitzEspositoPRX2014, 
KoskiEtAlPRL2015, StrasbergEtAlPRB2018, PtaszynskiEspositoPRL2019, SanchezSamuelssonPottsPRR2019} or the 
thermodynamics of various forms of information processing~\cite{MandalJarzynskiPNAS2012, DeffnerJarzynskiPRX2013, 
StrasbergEtAlPRE2015}. Of particular inspiration in our context is the approach by Deffner and 
Jarzynski~\cite{DeffnerJarzynskiPRX2013}, hence it is also worthwhile to distinguish our approach from it. First 
and most importantly, they did not explicitly connect their autonomous approach to an observer-dependent point of view 
to obtain the corresponding thermodynamic definitions at the trajectory level. Second and more an issue of 
technicalities, their approach was classical, used certain weak coupling assumptions, and they treated information and 
entropy differently by excluding correlations, which turn out to be crucial for our purposes. By overcoming all these 
assumptions, we do not only justify the framework of Refs.~\cite{StrasbergPRE2019, StrasbergWinterPRE2019, 
StrasbergPRL2019}, but we provide a general and promising tool to study the emergence of thermodynamic quantities at the 
trajectory level \emph{without} making explicit use of quantum measurements. This opens up a novel possibility 
to derive the laws of thermodynamics at the trajectory level even beyond the present considerations. 

A short summary together with an outline of the paper reads as follows. First, in 
Sec.~\ref{sec quantum causal models} we briefly review the essential of a quantum causal model or quantum stochastic 
process as far as it is needed for the following. Afterwards in Sec.~\ref{sec autonomous model}, we carefully 
construct the corresponding autonomous model, whose dynamical equivalence to a quantum causal model is proven in 
Sec.~\ref{sec dynamical equivalence}. The central part of this paper is Sec.~\ref{sec thermodynamic equivalence}. 
In there, we study the thermodynamics of our autonomous model using arguments from statistical mechanics and we 
demonstrate that it naturally induces definitions at the trajectory level in accordance with operational quantum 
stochastic thermodynamics (apart from one minor and typically negligible difference). Furthermore, we also 
discover that our autonomous approach leaves room for some remaining ambiguities in the definition of 
stochastic heat and work. In light of this freedom, we show that the definition of stochastic work in the `two-point 
projective measurement scheme'~\cite{EspositoHarbolaMukamelRMP2009, CampisiHaenggiTalknerRMP2011} provides one possible 
consistent choice within our autonomous approach. However, further arguments show that it is only valid for 
\emph{isolated}, but not for open systems (in which we are primarily interested here). On the other hand, the concept 
of ``quantum heat'', at least as originally introduced in Ref.~\cite{ElouardEtAlQInf2017}, does not have any theoretical 
foundation within our autonomous approach. Finally, the paper ends with some additional noteworthy remarks in 
Sec.~\ref{sec remarks}. 

%%%%%%%%%%%%%%%%%%%%%%%%%%%%%%%%%%%%%%%%%%%%%%%%%%%%%%%%%%%%%%%%%%%%%%%%%%%%%%%%%%%%%%%%%%%%%%%%%%%%%%%%%%%%%%%%%%%%%%%%
\section{Quantum causal models}
\label{sec quantum causal models}

Albeit there are some differences in the detailed mathematical description of a quantum causal model or quantum 
stochastic process~\cite{LindbladCMP1979, AccardiFrigerioLewis1982, ChiribellaDArianoPerinottiPRA2009, 
HardyRSocA2012, OreshkovCostaBruknerNC2012, CostaShrapnelNJP2016, OreshkovGiarmatziNJP2016, PollockEtAlPRA2018, 
MilzEtAlArXiv2017}, the common idea is that the primary entity in an experiment is the control operation or 
intervention performed on the system, but not the state (i.e., density operator) of the system itself. 
By shifting one level higher from states to operations, a quantum causal model can be represented by a multi-linear map 
from the set of interventions (applied at different times) to a final output state. While being quite abstract at first 
place, it offers many conceptual advantages, for instance, to optimize quantum 
circuits~\cite{ChiribellaDArianoPerinottiPRL2008, QuintinoEtAlPRA2019}, to rigorously define quantum 
non-Markovianity~\cite{PollockEtAlPRL2018} or classicality~\cite{StrasbergDiazPRA2019, MilzEgloffEtAlArXiv2019} in 
quantum processes, as well as to design multi-time resource theories~\cite{BerkEtAlArXiv2019}. If the system is 
classical, the approach reduces to classical causal modeling~\cite{PearlBook2009}, which allows to go beyond the 
standard description of a classical stochastic process, which is based only on passive and perfect observations. We 
here follow closely Refs.~\cite{PollockEtAlPRA2018, MilzEtAlArXiv2017}, which has a clear interpretation in terms of 
stochastic trajectories, see also Ref.~\cite{SakuldeeEtAlJPA2018}. We note that we do not attempt to review the 
approach in its full generality.

To begin with, we briefly repeat the essential of quantum operations, instruments and interventions~\cite{KrausBook1983, 
HolevoBook2001b}. At any time any such operation is described by a completely positive map $\C A(r)$, where $r$ denotes 
the measurement result associated to this intervention. This could be the result of a standard projective measurement or 
a more general measurement~\cite{WisemanMilburnBook2010, JacobsBook2014}. Its action on the density operator $\rho_S$ of 
the system is denoted as $\tilde\rho_S(r) = \C A(r)\rho_S$, where we used a `tilde' to denote a non-normalized state 
$\tilde\rho_S(r)$. The probability to obtain outcome $r$ is encoded in its trace $p(r) = \mbox{tr}_S\{\tilde\rho_S(r)\}$. 
A set of completely positive maps $\C A(r)$ forms an instrument if its average effect $\C A \equiv \sum_r\C A(r)$ is 
described by the completely positive and trace-preserving map $\C A$. It can be written in the familiar operator-sum 
representation $\C A\rho_S = \sum_i K_i\rho_S K_i^\dagger$ with $\sum_i K_i^\dagger K_i = 1_S$, where it is also known 
as a Kraus map. Here, $1_S$ denotes the identity in the system Hilbert space. 

By generalizing to multiple times, we now allow that the external agent interrupts the time-evolution of the 
system at arbitrary times $t_n > \dots > t_0$ by arbitrary interventions characterized by an instrument 
$\{\C A_k(r_k)\}$. Here, the subscripts indicate the time $t_k$ at which the intervention happens. Note that at each 
time $t_k$ we can choose a different instrument with, possibly, a different set of measurement results associated to it. 
Given a sequence of measurement results, denoted by $\bb r_n = (r_0,\dots,r_n)$, the non-normalized state of the system 
at a time $t>t_n$ can be formally written as 
\begin{equation}\label{eq process tensor}
 \tilde\rho_S(t,\bb r_n) = \mathfrak{T}[\C A_n(r_n),\dots,\C A_0(r_0)].
\end{equation}
Here, the so-called `process tensor'~\cite{PollockEtAlPRA2018, MilzEtAlArXiv2017, PollockEtAlPRL2018} 
$\mathfrak{T}$ presents a multi-linear map from the set of control operations to the final system state. 
Note that we only indicated the dependence on the control operations $\C A_k(r_k)$ because those are the objects 
we assume to be controllable in an experiment. In contrast, for instance, we did not indicate the dependence on the 
initial system state, which is assumed to be arbitrary but fixed [albeit it can be manipulated via $\C A_0(r_0)$]. 
In particular, in case of system-environment correlations the process tensor does not depend linearly on the initial 
system state. Furthermore, we remark that the process tensor can be tomographically reconstructed by measuring the 
final output state many times in response to the chosen set of control operations and hence, it is experimentally a 
well-defined object. Finally, the probability to get the results $\bb r_n$ is given via 
$p(\bb r_n) = \mbox{tr}_S\{\tilde\rho_S(t,\bb r_n)\}$. 

Microscopically, the process tensor arises from the following picture. Let 
\begin{equation}\label{eq Hamiltonian SB}
 H_{SB}(\lambda_t) = H_S(\lambda_t) + H_B + V_{SB}
\end{equation}
denote an arbitrary system-bath Hamiltonian, where $H_S$ ($H_B$) is the bare system (bath) Hamiltonian and $V_{SB}$ 
their mutual interaction. Furthermore, in view of the thermodynamic framework considered later on, we already introduced 
some time-dependent driving protocol $\lambda_t$ (e.g., an electric or magnetic field), which can change the energies 
of the system. For the present section, however, this is of minor relevance. The global unitary time evolution from $t$ 
to $t'$ is described by the superoperator 
\begin{equation}
 \begin{split}\label{eq unitary SB}
  \rho_{SB}(t') &= \C U_{SB}(t',t) \rho_{SB}(t) \\
  &\equiv U_{SB}(t',t) \rho_{SB}(t) U_{SB}^\dagger(t',t),
 \end{split}
\end{equation}
where $U_{SB}(t',t) = \C T_+\exp\left[-i\int_t^{t'} ds H_{SB}(\lambda_s)/\hbar\right]$ with the time-ordering 
operator $\C T_+$. Then, the process tensor can be microscopically expressed as 
\begin{equation}
 \begin{split}
  \tilde\rho_S(t,\bb r_n) =&~ \mathfrak{T}[\C A_n(r_n),\dots,\C A_0(r_0)] \label{eq process tensor microscopic} \\
  =&~ \mbox{tr}_B\left\{\C U_{SB}(t,t_n)\C A_n(r_n)\dots\right. \\
  &\left.~~~~~~\dots\C U_{SB}(t_1,t_0)\C A_0(r_0)\rho_{SB}(t_0^-)\right\}.
 \end{split}
\end{equation}
Here, $\rho_{SB}(t_0^-)$ denotes the global system-bath state, which can be arbitrary at the moment, prior to the 
first intervention, which happens at time $t_0$. Note that we use in general the notation $t^\pm$ to denote a point 
in time just before or after $t$. Furthermore, the multi-linearity of the process tensor is evident from 
Eq.~(\ref{eq process tensor microscopic}). Finally, remember that each $\C A_k(r_k) \equiv \C A_k(r_k)\otimes\C I_B$ 
acts non-trivially only on the system (we suppress any identity operations $\C I$ as well as many tensor products 
in the notation). 

It turns out~\cite{ChiribellaDArianoPerinottiPRA2009, HardyRSocA2012, OreshkovCostaBruknerNC2012, CostaShrapnelNJP2016, 
OreshkovGiarmatziNJP2016, PollockEtAlPRA2018, MilzEtAlArXiv2017} that the framework can be even further generalized. 
Remember that $\mf T$ is a process \emph{tensor} acting multi-linearly on a sequence of interventions. Equivalently, the 
process tensor can be seen as an object that acts on the tensor product of spaces $\C L(\C H_S\otimes\C H_S)$, where 
$\C L(\C H_S\otimes\C H_S)$ denotes the vector space of linear maps acting on $\C H_S\otimes\C H_S$ with the system 
Hilbert space $\C H_S$. Thus, if we denote by $\bb{A}_{n:0}(\bb r_n) \equiv \C A_n(r_n)\otimes\dots\otimes\C A_0(r_0)$ 
an element of that space, we can write Eq.~(\ref{eq process tensor}) in short as 
$\tilde\rho_S(t,\bb r_n) = \mf{T}\bb{A}_{n:0}(\bb r_n)$. Now, due to linearity, it is possible to consider any 
sequence of control operations $\bb A'$, not only those that are decorrelated as $\bb{A}_{n:0}(\bb r_n)$ is. 
This happens, for instance, when one considers the average effect of classical feedback control where 
$\C A_k(r_k) = \C A_k(r_k|\bb r_{k-1})$ depends on previous measurement results. Note that also the driving protocol 
$\lambda_t = \lambda_t(\bb r_{k-1})$ is allowed to depend on previous measurement results. This generality captures 
any conceivable feedback scenario, but for notational simplicity we suppress the possible dependence on $\bb r_{k-1}$ 
most of the times. Furthermore, it is even possible to consider quantum correlated operations. This goes beyond 
classical feedback control and can result in interventions that can no longer be written as a completely positive 
map at a single time (the overall process tensor nevertheless preserves complete positivity). It will become clear 
from the exposition below that we can also include this into our autonomous framework, but for ease of presentation we 
refrain from discussing the most general scenario with all its details. Finally, within the framework of quantum 
causal models it is even possible to consider interventions $\C A_k(r_k)$, where the input and output spaces are 
different (for instance, by adding or discarding ancillas to or from the system), or space-like separated interventions, 
which happen at different laboratories. Again, we find that the benefits added by the greater generality do not 
outweigh the drawbacks of a more hampered presentation here. 

%%%%%%%%%%%%%%%%%%%%%%%%%%%%%%%%%%%%%%%%%%%%%%%%%%%%%%%%%%%%%%%%%%%%%%%%%%%%%%%%%%%%%%%%%%%%%%%%%%%%%%%%%%%%%%%%%%%%%%%%
\section{Autonomous model}
\label{sec autonomous model}

\begin{figure}%[h]
 \centering\includegraphics[width=0.42\textwidth,clip=true]{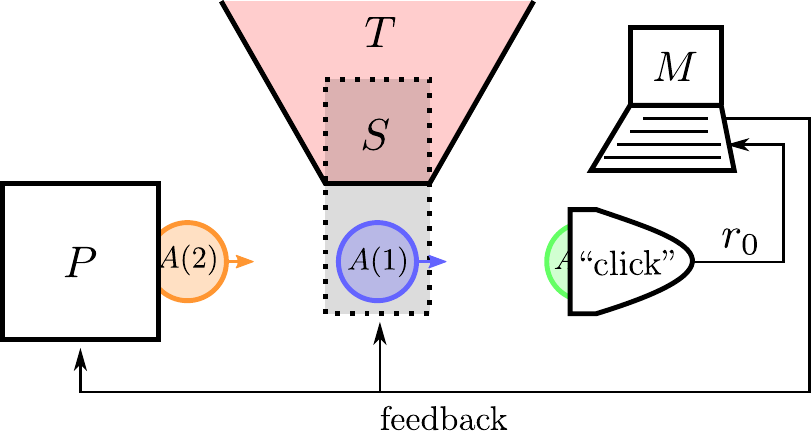}
 \label{fig setup} 
 \caption{A system $S$ is in contact with a bath, which -- in view of the thermodynamic framework considered later on -- 
 is sketched as a heat bath with initial temperature $T$. A preparation apparatus $P$ sequentially produces ancillas 
 $A(k)$, $k=0,1,\dots$, which interact with the system when they enter the shaded grey area and thereby implement a 
 control operation. Afterwards, these ancillas are detected giving rise to a measurement outcome $r_k$, which is stored 
 in a memory $M$. As indicated by the feedback loop, the external agent can decide to change, e.g., the state of each 
 ancilla (sketched with different colors) or the Hamiltonian of the system or the system-ancilla interaction via the 
 protocol $\lambda_t$ (not explicitly sketched) conditioned on all previous outcomes. }
\end{figure}

In this section we construct the autonomous model, which simulates a quantum stochastic process of the 
form~(\ref{eq process tensor}) if it is finally subjected to an appropriate measurement giving result $\bb r_n$. 
That this is in principle possible is not new, see Refs.~\cite{ChiribellaDArianoPerinottiPRA2009, SilvaEtAlNJP2017, 
PollockEtAlPRA2018}. Our discussion is, however, less abstract and more `physics'-oriented by explicitly specifying 
Hamiltonians. This is needed later on to formulate a theory of thermodynamics. We will guide our construction along the 
experimental setup sketched in Fig.~\ref{fig setup}. We proceed in two steps. 

First, we only consider the unconditional or unmeasured dynamics. This means that the external agent only 
deterministically implements control operations $\C A_k$ at time $t_k$ described by completely positive and 
trace-preserving maps, which do not depend on any measurement result $r_k$. Pictorially speaking, we ignore the right 
hand side of Fig.~\ref{fig setup} (the detector, the memory, and the feedback loop). Then, the main insight to get 
an autonomous Hamiltonian model for this situations rests on the unitary dilation theorem, first proven by 
Stinespring~\cite{StinespringPAMS1955} (see also Refs.~\cite{KrausBook1983, HolevoBook2001b}). It states that any 
control operation can be written as the reduced dynamics of a unitary interaction with an external ancilla system: 
\begin{equation}\label{eq Stinespring}
 \C A_k \rho_S = \mbox{tr}_{A(k)}\{U_{SA(k)}\rho_S\otimes\rho_{A(k)}U_{SA(k)}^\dagger\}.
\end{equation}
Here, $U_{SA(k)}$ denotes the unitary operator resulting from the system-ancilla interaction and $\rho_{A(k)}$ the 
initial state of the $k$th ancilla, which was prepared in a preparation apparatus $P$. Note that the unitary and 
the initial state are allowed to depend on $k$ such that, in general, $\C A_k \neq \C A_\ell$ for $k\neq\ell$. 
The Hamiltonian associated to this `unconditional' setup therefore reads 
\begin{equation}\label{eq Hamiltonian uncond}
 H_{SBPA}(\lambda_t) = H_{SB}(\lambda_t) + H_{PA}(\lambda_t) + H_{SA}(\lambda_t).
\end{equation}
In detail, it consists of the following parts: 

\bb{A. System-bath part $H_{SB}(\lambda_t)$.} This is the same as in Eq.~(\ref{eq Hamiltonian SB}) describing 
the system, bath and their interaction ignoring any external influence. 
 
\bb{B. Ancilla preparation $H_{PA}(\lambda_t)$.} In this part the different ancillas are produced by implementing 
a unitary $U_{PA(k)}$ prior to the interaction of ancilla $A(k)$ with the system. By fixing a suitable initial state 
$\rho_P(t_0^-)$ of the preparation apparatus, we can -- due to Eq.~(\ref{eq Stinespring}) by choosing an appropriate 
$U_{PA(k)}$ -- implement any operation we want on the ancilla. Hence, we can prepare any ancilla state we 
like~\cite{WuEtAlJPA2007}. Due to this, the initial state of the ancillas $\rho_{A}(t_0^-)$ can be in principle 
arbitrary, albeit in any experiment there are certain restrictions imposed on the preparation of the initial ancilla 
states, see, e.g., Refs.~\cite{SayrinEtAlNature2011, ZhouEtAlPRL2012}. Note that we use $A$ to denote the totality 
of all ancillas $A(0), A(1), \dots, A(n)$ and that $n$ can be an arbitrary large number. 
 
\bb{C. System-ancilla part $H_{SA}(\lambda_t)$.} This Hamiltonian reads in detail 
\begin{equation}\label{eq Hamiltonian SA}
 H_{SA}(\lambda_t) = \sum_{k=0}^n \left[H_{A(k)} + V_{SA(k)}(\lambda_t)\right]
\end{equation}
and describes the bare Hamiltonian $H_{A(k)}$ of each ancilla $A(k)$ as well as its interaction $V_{SA(k)}(\lambda_t)$ 
with the system. Each $H_{A(k)}$ can be different and in principle even time-dependent, albeit this is typically not 
the case and therefore, we omitted it for notational simplicity. In contrast, the time-dependence of 
$V_{SA(k)}(\lambda_t)$, which can be again different for each $A(k)$, is crucial. Later on in 
Sec.~\ref{sec dynamical equivalence} we will design it in such a way that it implements the unitary $U_{SA(k)}$ in 
Eq.~(\ref{eq Stinespring}). At the moment, however, we are more relaxed and only assume that $V_{SA(k)}(\lambda_t)$ is 
zero outside the `interaction zone' with the system (the shaded grey area in 
Fig.~\ref{fig setup}). Especially, it is zero when the ancilla gets prepared in $P$ or measured afterwards (see below). 
 
\bb{D. Work reservoir $\lambda_t$.} We still allow for an external time-dependent field $\lambda_t$, which is 
responsible for, e.g., changing the system Hamiltonian $H_S(\lambda_t)$ or switching on and off the system-ancilla 
interactions $V_{SA(k)}(\lambda_t)$. This means that we model the driving, which will be later on identified with 
the work supplied to the setup, semi-classically. While this is not fully autonomous (in the sense of a completely 
time-independent model), the resulting dynamics are nevertheless unitary. Note that the ideal limit needed to generate 
a time-dependent Hamiltonian out of a time-independent one is understood~\cite{DeffnerJarzynskiPRX2013}. As the 
purpose of this paper is not  to understand the detailed autonomous modeling of work reservoirs, we stick 
throughout to this semi-classical picture for ease of presentation. 

We remark that the setup specified so far is identical to the framework of repeated interactions or collisional models 
as considered in Refs.~\cite{BruneauJoyeMerkliAHP2010, StrasbergEtAlPRX2017, CresserPS2019, StrasbergPRL2019}. Next, we 
want to explicitly include measurements and conditioning in the description. Here, the key mathematical ingredient to 
autonomously model the observer is an extension of Eq.~(\ref{eq Stinespring}). In fact, every possible intervention 
$\C A_k(r_k)$ can be implemented as~\cite{HolevoBook2001b, OzawaJMP1984} 
\begin{equation}
 \begin{split}
  \C A& _k(r_k)\rho_S = \label{eq Stinespring instruments} \\
  &\mbox{tr}_{A(k)}\left\{P(r_k)U_{SA(k)}\rho_S\otimes\rho_{A(k)}U_{SA(k)}^\dagger\right\},
 \end{split}
\end{equation}
where $P(r_k)$ is some orthogonal resolution of the identity in the ancilla Hilbert space, 
$\sum_{r_k} P(r_k) = 1_{A(k)}$. Note that the average effect of the intervention~(\ref{eq Stinespring instruments}) 
is described by Eq.~(\ref{eq Stinespring}), i.e., $\sum_{r_k}\C A_k(r_k) = \C A_k$. To implement 
Eq.~(\ref{eq Stinespring instruments}), we need additional degrees of freedom. They will turn out to describe 
an idealized classical memory responsible for performing the measurement of the ancilla and for storing the measurement 
result $r_k$. Finally, we also need to implement the feedback loop as sketched in Fig.~\ref{fig setup} in an autonomous 
way, but this does not need any additional physical degrees of freedom. Thus, the 
Hamiltonian~(\ref{eq Hamiltonian uncond}) is generalized to
\begin{equation}
 \begin{split}\label{eq Hamiltonian tot}
  H_\text{tot}(\lambda_t) =&~ H_M(\lambda_t) + V_{AM}(\lambda_t) \\ 
                          &  + \sum_{\bb r_n} H_{SBPA}(\lambda_t,\bb r_n)|\bb r_n\rl\bb r_n|.
 \end{split}
\end{equation}
We now study its terms again separately in detail. 

\bb{E. Memory part $H_M(\lambda_t)$.} Following the tradition of the thermodynamics of 
computation~\cite{BennettIJTP1982}, we split the memory in informational degrees of freedom (IDF) $I$ and 
non-informational degrees of freedom (NIDF) $N$, which are here responsible for dephasing the IDF (see also 
Ref.~\cite{DeffnerJarzynskiPRX2013}). Strictly speaking, the NIDF are not necessary for the 
following, but we keep them as they simplify the algebra and argumentation at some places and, in particular, including 
them seems more realistic from a physical perspective. Thus, the Hamiltonian of the memory is split as 
\begin{equation}\label{eq Hamiltonian M}
 H_M(\lambda_t) = H_I + H_N + V_{IN}(\lambda_t). 
\end{equation}
The Hilbert space of the IDF is spanned by the vectors $|\bb r_n\rangle = |r_n\rangle\otimes\dots\otimes|r_0\rangle$ 
encoding the measurement results. As customarily done, we assume that these states are energetically degenerate, 
i.e., $H_I \sim 1_I$. Furthermore, the IDF are initially in a standard reference state 
$\rho_I(t_0^-) = |\bb 1_n\rl\bb 1_n| = |1\rl1|\otimes\dots\otimes|1\rl1|$ decorrelated from the rest. 
We assume that the NIDF act like a pure dephasing bath such that the information stored in $I$ is classical meaning 
that, after tracing out the NIDF, the IDF are only classically correlated with the rest: 
\begin{equation}
 \begin{split}
  \rho_{SBPAI}(t) &= \sum_{\bb r_n} \tilde\rho_{SBPA}(t,\bb r_n) \otimes |\bb r_n\rl\bb r_n| \label{eq state BSAI} \\
                  &= \sum_{\bb r_n} p(\bb r_n)\rho_{SBPA}(t,\bb r_n) \otimes |\bb r_n\rl\bb r_n|.
 \end{split}
\end{equation}
The dephasing can be implemented in various ways and, in principle, does not entail any energetic cost. An explicit 
example works as follows\footnote{This part can be skipped by readers, who know how to implement a dephasing 
operation in a unitary way without energy cost.}: let $r_k\in\{1,\dots,d(k)\}$ label the in total $d(k)$ different 
measurement results at time $t_k$. Then, let $H_N$ describe a set of $n$ non-interacting and energetically degenerate 
entities, which are prepared in a maximally mixed state of dimension $d(k)$ respectively, 
$\rho_N(t_0^-) = 1_{d(0)}/d(0)\otimes\dots\otimes 1_{d(n)}/d(n)$ (note that a maximally mixed state is identical 
to a Gibbs state for degenerate energies). Then, let $V_{IN}(\lambda_t)$ implement a short unitary evolution between 
$I(k)$ and $N(k)$, which happens right after the $k$th measurement and has the form 
$U_{IN(k)} = \sum_{r_k}|r_k\rl r_k| \otimes \sum_{i=1}^{d(k)}|i+r_k\rl i|$ [we interpret $|i+r_k\rangle$ modulo $d(k)$ 
if $i+r_k > d(k)$]. 
Due to the degeneracy it is obvious that $[U_{IN(k)},H_I + H_N] = 0$ 
and thus, the unitary has no energetic cost. Furthermore, straightforward algebra shows that 
\begin{equation}
 \begin{split}\label{eq dephasing}
  \C D^{(k)}\rho_I 
  &\equiv \mbox{tr}_N\left\{U_{IN(k)}\rho_I\otimes\rho_N(t_0^-)U_{IN(k)}^\dagger\right\} \\ 
  &= \sum_{r_k} |r_k\rl r_k|\rho_I|r_k\rl r_k|.
 \end{split}
\end{equation}
Thus, we implemented a dephasing operation at zero energetic cost, as desired. 
 
\bb{F. Ancilla-memory part $V_{AM}(\lambda_t)$.} This part is responsible for the actual measurement of the 
ancilla by correlating its state with the IDF, i.e., 
$V_{AM}(\lambda_t) = V_{AI}(\lambda_t) = \sum_{k=0}^n V_{AI(k)}(\lambda_t)$, where we assumed that the $k$th IDF is 
responsible for the measurement of ancilla $A(k)$. The desired unitary reads 
$U_{AI(k)} = \sum_{r_k} P(r_k) \otimes\sum_{i}|i+r_k-1\rl i|_{I(k)}$ such that for any $\rho'_{A(k)}$ 
[we use a primed notation to distinguish it from the state $\rho_{A(k)}$ appearing in 
Eqs.~(\ref{eq Stinespring}) and~(\ref{eq Stinespring instruments})]
\begin{equation}
 \begin{split}\label{eq measurement ancilla}
  U_{AI(k)}&\rho'_{A(k)}|1\rl1| U^\dagger_{AI(k)} = \\
  &\sum_{r_k,r'_k} P(r_k)\rho'_{A(k)} P(r'_k)|r_k\rl r'_k|.
 \end{split}
\end{equation}
Thus, if we measure the IDF in state $|r_k\rangle$, the conditional state of the ancilla is 
$P(r_k)\rho'_{A(k)} P(r_k)$, which eventually gives rise to Eq.~(\ref{eq Stinespring instruments}).
Note that the time-dependence of $V_{AI(k)}(\lambda_t)$ is such that the measurement happens 
\emph{after} the interaction between the system and the $k$th ancilla as implemented by Eq.~(\ref{eq Hamiltonian SA}), 
but \emph{before} the dephasing operation~(\ref{eq dephasing}). 
 
\bb{G. Conditional (feedback) part.} So far, the external agent can implement arbitrary 
control operations $\C A_k(r_k)$ at an arbitrary set of discrete times $t_k$. However, in the most general case, the 
external agent is also allowed to use the available information in the memory to condition the future dynamics after 
time $t>t_k$ on the so far available measurement results $\bb r_k$. This is implemented by the last part of 
Eq.~(\ref{eq Hamiltonian tot}), $\sum_{\bb r_n} H_{SBPA}(\lambda_t,\bb r_n)|\bb r_n\rl\bb r_n|$, which applies a 
different `unconditional' Hamiltonian~(\ref{eq Hamiltonian uncond}) depending on the state of the memory 
$|\bb r_n\rl\bb r_n|$. In fact, due to Eq.~(\ref{eq state BSAI}) the evolution from time $t_k^+$ to $t_{k+1}^-$ 
is given by 
\begin{align}
 &\rho_{SBPAI}(t_{k+1}^-) = \\
 &~~ \sum_{\bb r_k} U_{SBPA}(\bb r_k)\rho_{SBPA}(\bb r_k,t_k^+)U_{SBPA}^\dagger(\bb r_k) |\bb r_k\rl\bb r_k|, \nonumber
\end{align}
where 
\begin{equation}
 \begin{split}
  U&_{SBPA}(\bb r_k) =\\ & \C T_+\exp\left[-i\int_{t_k}^{t_{k+1}} H_{SBPA}(\lambda_s,\bb r_k) ds \right].
 \end{split}
\end{equation}
Here, we excluded the results $r_\ell$ for $\ell>k$ because we naturally assume that for $t<t_\ell$ 
$H_{SBPA}(\lambda_t,\bb r_n) = H_{SBPA}(\lambda_t,\bb r_{\ell-1})$ depends only on the so far obtained measurement 
results. To conclude, for each measurement trajectory $\bb r_k$ we can apply a different Hamiltonian affecting any 
possible part of Eq.~(\ref{eq Hamiltonian uncond}) and hence, allowing full control about the system and the ancillas. 
If we do not perform feedback, then $H_{SBPA}(\lambda_t,\bb r_n) = H_{SBPA}(\lambda_t)$ for all $\bb r_n$. Note that 
we could even change the time of the measurements during the experiment by conditioning the memory Hamiltonian 
$H_M(\lambda_t)$ on previous measurement results too. For ease of presentation we refrained from writing down the most 
general case. Finally, we remark that the present construction can be seen as a general form of coherent 
feedback control~\cite{WisemanPhD1994, WisemanMilburnPRA1994b, LloydPRA2000}. It was already used to study the 
thermodynamics of feedback control in Refs.~\cite{JacobsPRA2009, StrasbergEtAlPRX2017}. 

We repeat that the temporal order of the dynamics is essential (see also Fig.~\ref{fig setup time}): the 
preparation happens before the actual control operation (the system-ancilla interaction), which happens before the 
measurement of the ancilla, which happens before the final dephasing of the memory. Apart from this order the 
time-dependence of all interactions is so far arbitrary. 

Finally, the time evolution is fully fixed by specifying the global initial state, which reads 
\begin{equation}\label{eq initial state}
 \rho_\text{tot}(t_0^-) = \rho_{SB}(t_0^-)\rho_P(t_0^-)\rho_A(t_0^-)\rho_M(t_0^-).
\end{equation}
Here, $\rho_{SB}(t_0^-)$ is an arbitrary initial system-bath state, which we will need to restrict in 
Sec.~\ref{sec thermodynamic equivalence}, $\rho_P(t_0^-)$ is a suitable chosen initial state of the preparation 
apparatus, $\rho_A(t_0^-)$ is an arbitrary initial ancilla state, and finally, the initial state of the memory is chosen 
as $\rho_M(t_0^-) = |\bb 1_n\rl\bb 1_n|_I\rho_{N}(t_0^-)$ with a suitable initial state for the NIDF as discussed above. 

%%%%%%%%%%%%%%%%%%%%%%%%%%%%%%%%%%%%%%%%%%%%%%%%%%%%%%%%%%%%%%%%%%%%%%%%%%%%%%%%%%%%%%%%%%%%%%%%%%%%%%%%%%%%%%%%%%%%%%%%
\section{Dynamical equivalence with a quantum causal model}
\label{sec dynamical equivalence}

We now show that our autonomous model captures the dynamics of a quantum causal model as described in 
Sec.~\ref{sec quantum causal models}. For that purpose we need to implement the control operations instantaneously. 
Ideally, this requires that the interaction between the system and the $k$th ancilla can be written as 
\begin{equation}\label{eq instantaneous interaction}
 V_{SA(k)}(\lambda_t) = \delta(t-t_k)\ln(iU_{SA(k)}),
\end{equation}
where $\delta(t-t_k)$ denotes the Dirac delta function. This implements an instantaneous unitary evolution 
$U_{SA(k)}$ at time $t_k$. 

\begin{figure}%[b]
 \centering\includegraphics[width=0.44\textwidth,clip=true]{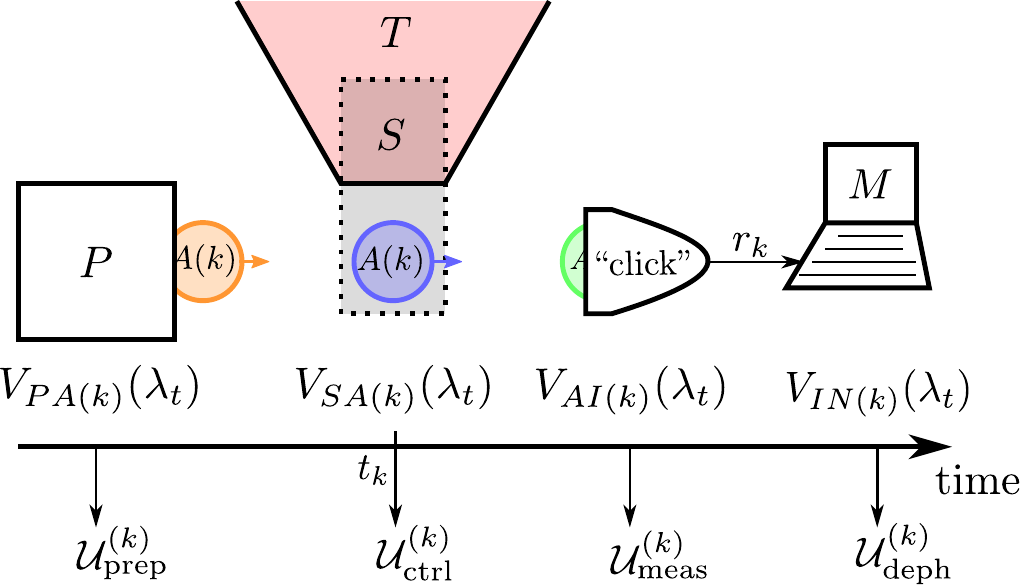}
 \label{fig setup time} 
 \caption{This figure illustrates the temporal order of the setup, in which a control operation is 
 implemented within our autonomous setup. First, the $k$th ancilla gets prepared due to the interaction 
 $V_{PA(k)}(\lambda_t)$ ultimately resulting in the operation $\C U_\text{prep}^{(k)}$ (see 
 Sec.~\ref{sec dynamical equivalence}). Then, the ancilla interacts with the system via $V_{SA(k)}(\lambda_t)$ and, in 
 the limit where this interactions happens instantaneously (also see Sec.~\ref{sec dynamical equivalence}), creates the 
 operation $\C U_\text{ctrl}^{(k)}$. Afterwards, the ancilla gets `detected' via the interaction $V_{AI(k)}(\lambda_t)$ 
 with the IDF as described by the operation $\C U_\text{meas}^{(k)}$. Finally, the NIDF dephase the IDF via 
 $V_{IN(k)}(\lambda_t)$ in turn creating the operation $\C U_\text{deph}^{(k)}$. We remark that there is some freedom 
 of how to fix the time $t_k$, when the intervention `happens'. Here, it is indicated as the time when the system-ancilla 
 interaction takes place, which is well-defined in the limit where this interaction is instantaneous as assumed in 
 Sec.~\ref{sec dynamical equivalence}. Note that we excluded the feedback loop from Fig.~\ref{fig setup} for a simplified 
 graphical presentation only. }
\end{figure}

Starting from the initial state~(\ref{eq initial state}), the time evolution of the global state can be iteratively 
constructed via 
\begin{equation}\label{eq full unitary}
 \rho_\text{tot}(t_{k+1}^-) = \C U_{SB}^{(k)}\C U^{(k)}_\text{deph}\C U^{(k)}_\text{meas}\C U^{(k)}_\text{ctrl}\C U^{(k)}_\text{prep}\rho_\text{tot}(t_k^-).
\end{equation}
Here, $\C U_{SB}^{(k)}$ is the unitary system-bath evolution from $t_k$ to $t_{k+1}$ [cf.~Eq.~(\ref{eq unitary SB})], 
and $\C U^{(k)}_\text{prep}$, $\C U^{(k)}_\text{ctrl}$, $\C U^{(k)}_\text{meas}$, and $\C U^{(k)}_\text{deph}$ denote 
the operations resulting from the preparation of the $k$th ancilla, its interaction with the system, its measurement, 
and the final dephasing of the memory, respectively (see also Fig.~\ref{fig setup time}). While their temporal 
order is important, it is not necessary that $\C U^{(k)}_\text{deph}$, $\C U^{(k)}_\text{meas}$ or 
$\C U^{(k)}_\text{prep}$ happen instantaneously before or after the control operation $\C U^{(k)}_\text{ctrl}$ since 
they commute with $\C U_{SB}^{(k)}$. In fact, in an actual experiment delays are unavoidable and preparations and 
measurements can take a finite time~\cite{SayrinEtAlNature2011, ZhouEtAlPRL2012}. 

After tracing out the NIDF as well as all ancillas, which are no longer 
participating in the interaction and which we denote by $A_\text{out}$, we write Eq.~(\ref{eq full unitary}) as 
\begin{equation}
 \begin{split}
  \mbox{tr}&_{A_\text{out}N}\{\rho_\text{tot}(t_{k+1}^-)\} = \\
  & \C U_{SB}^{(k)}\C D^{(k)}\C U^{(k)}_\text{meas}\C U^{(k)}_\text{ctrl}\C U^{(k)}_\text{prep}
  \rho_{SBPA(k)}(t_k^-)
 \end{split}
\end{equation}
Notice that we have replaced $\C U^{(k)}_\text{deph}$ by the dephasing map~(\ref{eq dephasing}) and due to 
Eq.~(\ref{eq state BSAI}) we have 
\begin{align}
 & \rho_{SBPA(k)}(t_k^-) = \\ 
 & ~~~\sum_{\bb r_{k-1}}\tilde\rho_{SBPA(k)}(t_k^-,\bb r_{k-1}) \otimes|1,\bb r_{k-1}\rl1,\bb r_{k-1}|. \nonumber
\end{align}
Here, $|1,\bb r_{k-1}\rangle$ describes the state of the IDF before the measurement, where the 
$k$th register is still set to its standard state `1'. Furthermore, we assumed that only ancilla $A(k)$ is 
participating in the $k$th interaction, in principle more general scenarios are conceivable.\footnote{For instance, 
the present framework also allows to `recycle' an old ancilla and to let it interact again with the system. 
This could implement a quantum correlated operation as mentioned at the end of Sec.~\ref{sec quantum causal models}. 
For ease of presentation we refrain from discussing the most general scenario with all its details.} 
Now, we use the preparation apparatus $P$ to prepare any ancilla state $\rho_{A(k)} = \rho_{A(k)}(\bb r_{k-1})$ we like 
using $\C U^{(k)}_\text{prep}$. Due to Eq.~(\ref{eq Hamiltonian tot}) this preparation procedure is allowed to depend 
on the previous measurement results $\bb r_{k-1}$, which we typically suppress in the notation. After tracing out 
$P$, we get 
\begin{equation}
 \begin{split}
  \mbox{tr}&_{PA_\text{out}N}\{\rho_\text{tot}(t_{k+1}^-)\} 
  = \C U_{SB}^{(k)}\C D^{(k)}\C U^{(k)}_\text{meas}\C U^{(k)}_\text{ctrl} \\
  &\times\sum_{\bb r_{k-1}} \tilde\rho_{SB}(t_k^-,\bb r_{k-1})\rho_{A(k)}|1,\bb r_{k-1}\rl1,\bb r_{k-1}|.
 \end{split}
\end{equation}
Next, due to Eq.~(\ref{eq measurement ancilla}) the action of the ancilla measurement reads explicitly 
\begin{align}
 &\mbox{tr}_{PA_\text{out}N}\{\rho_\text{tot}(t_{k+1}^-)\} = \sum_{\bb r_k,r'_k}\C U_{SB}^{(k)}\C D^{(k)} \label{eq cat state} \\
 &~\times P(r_k) \left[\C U^{(k)}_\text{ctrl}\tilde\rho_{SB}(t_k^-,\bb r_{k-1})\rho_{A(k)}\right] P(r'_k)|\bb r_k\rl r'_k,\bb r_{k-1}|. \nonumber
\end{align}
Note that $\C U^{(k)}_\text{ctrl} = \C U^{(k)}_\text{ctrl}(\bb r_{k-1})$ can be conditioned on all previous measurement 
results due to Eq.~(\ref{eq Hamiltonian tot}). The second line of Eq.~(\ref{eq cat state}) describes a giant 
Schr\"odinger cat state with respect to the different superpositions of the measurement results $r_k$. This cat is 
killed by the dephasing operation: 
\begin{align}
 \mbox{tr}&_{PA_\text{out}N}\{\rho_\text{tot}(t_{k+1}^-)\} = \label{eq final autonomous state} \\
 & \C U_{SB}^{(k)}\sum_{\bb r_k} \C P(r_k) \left[\C U^{(k)}_\text{ctrl}\tilde\rho_{SB}(t_k^-,\bb r_{k-1})\rho_{A(k)}\right] |\bb r_k\rl\bb r_k|, \nonumber
\end{align}
where we introduced the superoperator $\C P(r_k)\rho_{A(k)} \equiv P(r_k)\rho_{A(k)}P(r_k)$ corresponding to the 
measurement result $r_k$. 

Equation~(\ref{eq final autonomous state}) describes the state of the system, the bath, the $k$th ancilla, and 
the IDF of our autonomous black box model at the $k$th time step. To verify its equivalence with a quantum causal 
model, we imagine an external `super-observer' (who has engineered the black box), who reads out the IDF by performing 
a projective measurement. If the super-observer finds the results $\bb r_k$, the (non-normalized) conditional state of 
the bath, system and $k$th ancilla of the black box is according to Eq.~(\ref{eq final autonomous state}) 
\begin{equation}
 \begin{split}
  \tilde\rho_{SBA(k)}&(t_{k+1}^-,\bb r_k) = \\
  &\C U_{SB}^{(k)} \C P(r_k) \C U_\text{ctrl}^{(k)} \tilde\rho_{SB}(t_k^-,\bb r_{k-1}) \rho_{A(k)}.
 \end{split}
\end{equation}
After tracing out the bath and the ancilla and using Eq.~(\ref{eq Stinespring instruments}), we are left with 
\begin{equation}
 \tilde\rho_S(t_{k+1}^-,\bb r_k) = 
 \mbox{tr}_B\left\{\C U_{SB}^{(k)} \C A_k(r_k) \tilde\rho_{SB}(t_k^-,\bb r_{k-1})\right\}.
\end{equation}
If we iterate this, we arrive at Eq.~(\ref{eq process tensor microscopic}). This shows that our autonomous setup 
\emph{conditioned} on obtaining the measurement results $\bb r_k$ simulates any quantum causal model 
as introduced in Sec.~\ref{sec quantum causal models}. 

%%%%%%%%%%%%%%%%%%%%%%%%%%%%%%%%%%%%%%%%%%%%%%%%%%%%%%%%%%%%%%%%%%%%%%%%%%%%%%%%%%%%%%%%%%%%%%%%%%%%%%%%%%%%%%%%%%%%%%%%
\section{Thermodynamic equivalence with the operational framework}
\label{sec thermodynamic equivalence}

In this central section we derive thermodynamic definitions at the `unmeasured' level for our autonomous black box 
model (Sec.~\ref{sec thermo unmeasured}) and show that they naturally imply corresponding thermodynamic definitions at 
the trajectory level, which coincide with the definitions of Refs.~\cite{StrasbergPRE2019, StrasbergWinterPRE2019, 
StrasbergPRL2019} apart from one minor exception (Sec.~\ref{sec canonical choice}). However, we also discuss possible 
ambiguities at the stochastic level (Sec.~\ref{sec ambiguities}) and reconsider two other choices in the literature in 
light of our findings (Sec.~\ref{sec thermo other choices}). We start with some agreements though. 

%%%%%%%%%%%%%%%%%%%%%%%%%%%%%%%%%%%%%%%%%%%%%%%%%%%%%%%%%%%%%%%%%%%%%%%%%%%%%%%%%%%%%%%%%%%%%%%%%%%%%%%%%%%%%%%%%%%%%%%%
\subsection{Agreements}
\label{sec agreements}

The observer-dependent thermodynamic framework of Refs.~\cite{StrasbergPRE2019, StrasbergWinterPRE2019, StrasbergPRL2019} 
was derived under certain idealized assumptions, which we summarize here: 

\bb{I. Initial state.} The global form of the initial state~(\ref{eq initial state}) remains, but we assume 
that the initial system-bath state is described by a Gibbs ensemble denoted by $\pi$, i.e., 
$\rho_{SB}(t_0^-) = \pi_{SB}(\lambda_0) \equiv e^{-\beta H_{SB}(\lambda_0)}/\C Z_{SB}(\lambda_0)$. Note that this is 
in general a correlated state. We also assume that the NIDF are initially described by 
a Gibbs state as specified in Point~E above. They are initially decorrelated from the rest. 
 
\bb{II. Classical and fast memory.} The IDF are treated as an ideal classical memory. This implies that the 
IDF quickly dephase and, for all practically relevant times, are only classically correlated with the system and the 
ancillas, see Eq.~(\ref{eq state BSAI}). Furthermore, as already specified in Point~E above, the IDF are energetically 
degenerate and the dephasing operation is implemented without energetic cost. Finally, the measurement of the ancilla 
modeled by the interaction $V_{AI}(\lambda_t)$ is idealized to be infinitely fast, i.e., of the form 
$V_{AI}(\lambda_t) = \sum_k \delta(t-t'_k) \ln(iU_{AI(k)})$, where $t'_k$ denotes some time after the 
system-ancilla interaction. 
 
\bb{III. Preparation apparatus.} In principle, the preparation of the ancillas can have a thermodynamic cost. However, 
the goal of the repeated interaction framework is to include ancillas in an arbitrary nonequilibrium state into a 
consistent thermodynamic framework, regardless of how they were prepared~\cite{BruneauJoyeMerkliAHP2010, 
StrasbergEtAlPRX2017, CresserPS2019}. Consequently, also Refs.~\cite{StrasbergPRE2019, StrasbergWinterPRE2019, 
StrasbergPRL2019} ignored the preparation costs of the ancillas. In our context, it suffices to point out that, at 
least in principle, it is possible that the preparation has zero thermodynamic cost (for instance, by implementing 
the preparation reversibly). Since we are not interested in practical realization of our autonomous model, but rather 
in the theoretical foundations of quantum stochastic thermodynamics, we neglect in the following any discussion about 
the thermodynamic cost of the preparation and simply assume that it provides us with the desired ancillas. 

Finally, in this section we do \emph{not} assume that the system-ancilla interaction $V_{SA(k)}(\lambda_t)$ happens 
instantaneously, but it can instead take a finite time as also considered in Refs.~\cite{StrasbergEtAlPRX2017, 
StrasbergPRL2019}. In this sense we are more general here than in Sec.~\ref{sec dynamical equivalence}. Indeed, we 
discuss at the end that an instantaneous, delta-like interaction causes a subtle difference in the thermodynamic 
description. 

%%%%%%%%%%%%%%%%%%%%%%%%%%%%%%%%%%%%%%%%%%%%%%%%%%%%%%%%%%%%%%%%%%%%%%%%%%%%%%%%%%%%%%%%%%%%%%%%%%%%%%%%%%%%%%%%%%%%%%%%
\subsection{Thermodynamics at the unmeasured level}
\label{sec thermo unmeasured}

Our autonomous setup describes one big `supersystem' $SAI$, which consists of the system $S$, the ancillas $A$ and the 
IDF $I$, and which we label for the moment as $X = SAI$. It is connected to two heat baths: first, the bath $B$ in 
direct contact with the system $S$ and second, the NIDF $N$ responsible for dephasing the memory. The overall setup can 
therefore be recast in form of the Hamiltonian 
$H_\text{tot}(\lambda_t) = H_X(\lambda_t) + H_B + H_N + V_{XB} + V_{XN}(\lambda_t)$. The following results are based 
on two recent advances in strong coupling thermodynamics. First, we use the quantum 
version~\cite{StrasbergEspositoPRE2019} of the `Hamiltonian of mean force' framework~\cite{SeifertPRL2016} (see also 
Refs.~\cite{JarzynskiPRX2017, MillerAndersPRE2017, StrasbergEspositoPRE2017, HsiangHuEntropy2018, RivasArXiv2019} for 
related research in this direction). Then, we combine it with the framework 
of Refs.~\cite{EspositoLindenbergVandenBroeckNJP2010, TakaraHasegawaDriebePLA2010} to take into account the initially 
decorrelated dephasing bath. A detailed calculation how to combine the two frameworks can be found in the Supplement 
of Ref.~\cite{StrasbergPRL2019} and therefore we here only present its essential elements. 

We start with the conventional definition of mechanical work, which quantifies the global change in 
internal energy, i.e., 
$W(t)=\mbox{tr}\{H_\text{tot}(\lambda_t)\rho_\text{tot}(t)\}-\mbox{tr}\{H_\text{tot}(\lambda_0)\rho_\text{tot}(t_0)\}$. Furthermore, by construction the interaction $V_{XN}(\lambda_t)$ caused by the dephasing bath does not have 
any overall work cost, see Point~E above. Therefore, we can identify the total work with the work done on the 
supersystem $X$. It can be expressed as as an integral over the instantaneous 
supplied power: 
\begin{equation}\label{eq work X}
 W(t) = \int_{t_0}^t ds \mbox{tr}\left\{\frac{\partial H_X(\lambda_s)}{\partial s}\rho_X(s)\right\}.
\end{equation}
Note that, whenever it will be clear from context, we will suppress the subscript on the trace operation in the 
following. 

Next, we turn to the internal energy. To define it, we need the concept of the Hamiltonian of mean force, which is 
defined via the reduced equilibrium state of a global canonical Gibbs state. Specifically, with respect to an arbitrary 
system $X$ coupled to the bath $B$ we define 
\begin{equation}
 \pi_X^* \equiv \mbox{tr}_B\{\pi_{XB}\} \equiv \frac{e^{-\beta H_X^*}}{\C Z_X^*}, ~~~ 
 \C Z_X^* \equiv \frac{\C Z_{XB}}{\C Z_X}.
\end{equation}
This implicitly defines the Hamiltonian of mean force $H_X^*$. Note that $\pi_X^*\neq\pi_X$ in general. In addition, 
$H_X^*$ depends on the inverse temperature $\beta$ and the control parameter $\lambda_t$. Classically, it can 
be seen as an effective free energy landscape for the system, which is different from the bare energy $H_X$ due to the 
strong system-bath coupling. For readers unfamiliar with the framework of strong coupling thermodynamics, it might be 
easier to follow the rest of the paper by replacing the Hamiltonian of mean force $H_X^*$ with the standard Hamiltonian $H_X$, which amounts to assuming a weakly coupled heat bath. In fact, the main contribution of this paper is to provide 
a recipe to deduce trajectory-dependent thermodynamic definition from an autonomous picture without explicit 
measurements. With which thermodynamic definitions one starts at the unmeasured level is of rather minor relevance here. 
We only choose the strong coupling approach for the sake of generality to make clear that the resulting framework of 
operational quantum stochastic thermodynamics does not rely on the commonly used weak coupling or Markovian 
approximations.

We now define the internal energy of $X$ as 
\begin{equation}
 \begin{split}\label{eq internal energy pre}
  U(t) \equiv&~ 
  \mbox{tr}\left\{(H_X^* + \beta\partial_\beta H_X^*)\rho_X(t)\right\} \\
  &+ \mbox{tr}\{V_{XN}(\lambda_t)\rho_{XN}(t)\},
 \end{split}
\end{equation}
where $\partial_\beta$ denotes a partial derivative with respect to the inverse temperature. The first line coincides 
with the standard definition within the Hamiltonian of mean force framework~\cite{SeifertPRL2016, 
StrasbergEspositoPRE2019} and describes deviations from the weak coupling definition given by 
$\mbox{tr}\{H_X\rho_X(t)\}$. The second line in Eq.~(\ref{eq internal energy pre}) needs to be added to take 
into account the initially decoupled second bath~\cite{EspositoLindenbergVandenBroeckNJP2010, StrasbergPRL2019}. 
However, this expression can be simplified since the interaction $V_{XN}(\lambda_t) = V_{IN}(\lambda_t)$ responsible 
for the dephasing of the IDF is expected to act only very shortly after each measurement and hence, for practically all 
times we can set $\mbox{tr}\{V_{XN}(\lambda_t)\rho_{XN}(t)\} = 0$.\footnote{Alternatively, one could imagine a 
permanently but weakly coupled dephasing bath. Then, it also follows that 
$\mbox{tr}\{V_{XN}(\lambda_t)\rho_{XN}(t)\} \approx 0$.} Hence, 
\begin{equation}
 U(t) = \mbox{tr}\left\{(H_X^* + \beta\partial_\beta H_X^*)\rho_X(t)\right\}.
\end{equation}
Because we now  have a definition for work and internal energy, this automatically fixes the heat via the first law 
\begin{equation}\label{eq Q}
 Q(t) \equiv \Delta U(t) - W(t),
\end{equation}
where $\Delta U(t) = U(t) - U(t_0^-)$ denotes the change in internal energy. 

Let us now turn to the second law. First, we define the thermodynamic entropy of the supersystem $X$ 
\begin{equation}\label{eq S}
 S(t) \equiv S_\text{vN}[\rho_X(t)] + \beta^2\mbox{tr}\{(\partial_\beta H_X^*)\rho_X(t)\}.
\end{equation}
Here, $S_\text{vN}(\rho) \equiv -\mbox{tr}\{\rho\ln\rho\}$ denotes the von Neumann entropy and the second term is again 
a strong coupling correction~\cite{SeifertPRL2016, StrasbergEspositoPRE2019}. Then, the second law of nonequilibrium 
thermodynamics states that the entropy production $\Sigma$ is always positive, which can be expressed as ($k_B\equiv1$) 
\begin{equation}\label{eq 2nd law}
 \begin{split}
  \Sigma(t) &= \Delta S(t) - \beta Q(t) \\
            &= \beta[W(t)-\Delta F(t)] \ge 0.
 \end{split}
\end{equation}
Here, we defined the nonequilibrium free energy 
\begin{equation}\label{eq free energy}
 \begin{split}
  F(t)  &\equiv U - TS \\ 
        &= \mbox{tr}_X\{H_X^*\rho_X(t)\} - T S_\text{vN}[\rho_X(t)].
 \end{split}
\end{equation}
It differs from the conventional weak coupling definition solely by the replacement of $H_X$ with $H_X^*$.
The positivity of entropy production follows from monotonicity of relative entropy~\cite{UhlmannCMP1977, 
OhyaPetzBook1993} since 
\begin{equation}\label{eq 2nd law relative entropies}
 \begin{split}
  \Sigma(t) =&~ D[\rho_\text{tot}(t)\|\pi_{XB}(\lambda_t)\otimes\pi_N] \\
  &- D[\rho_X(t)\|\pi_X^*(\lambda_t)],
 \end{split}
\end{equation}
where $D[\rho\|\sigma] \equiv \mbox{tr}\{\rho(\ln\rho-\ln\sigma)\}$ denotes the quantum relative entropy. Showing the 
equivalence of Eqs.~(\ref{eq 2nd law}) and~(\ref{eq 2nd law relative entropies}) is tedious, but follows only 
standard steps, see the Supplement of Ref.~\cite{StrasbergPRL2019}. 

We now investigate the definitions above in detail by making extensive use of Eq.~(\ref{eq state BSAI}). 
First, the work~(\ref{eq work X}) originates from the three time-dependent terms $H_S(\lambda_t)$, $V_{SA}(\lambda_t)$, 
and $V_{AI}(\lambda_t)$. The first two contributions can be written as 
\begin{align}
 & W_S(t) = \label{eq work S} \\ 
  &~~~ \sum_{\bb r_n} p(\bb r_n) \int_{t_0}^t ds \mbox{tr}\left\{\frac{\partial H_S(\lambda_s,\bb r_n)}{\partial s}\rho_S(s,\bb r_n)\right\}, \nonumber \\
 &W_{SA}(t) = \label{eq work SA} \\ 
 &~~~ \sum_{\bb r_n} p(\bb r_n) \int_{t_0}^t ds 
 \mbox{tr}\left\{\frac{\partial V_{SA}(\lambda_s,\bb r_n)}{\partial s}\rho_{SA}(s,\bb r_n)\right\}. \nonumber
\end{align}
The third contribution due to $V_{AI}(\lambda_t)$ can be simplified by noting that the ancilla and IDF are 
isolated during the measurement such that we simply have to add up the changes in the ancilla energies (remember that 
the IDF are energetically degenerate). Thus, let $\rho'_{A(k)}(\bb r_{k-1})$ denote the state of the $k$th ancilla 
after the interaction with the system but before the measurement (which can depend on $\bb r_{k-1}$) and let 
$\rho''_{A(k)}(\bb r_k)$ denote its state after the measurement conditioned on finding the IDF in state $|r_k\rangle$. 
Then, if we split the work $W_{AI}(t) = \sum_k W_{AI(k)}(t)$ into its contributions due to the $k$th control step, 
we find that 
\begin{align}
 & W_{AI(k)}(t) = \label{eq work AI} \\
 & ~~\sum_{\bb r_k} p(\bb r_k) 
 \mbox{tr}\left\{H_{A(k)}[\rho''_{A(k)}(\bb r_k) - \rho'_{A(k)}(\bb r_{k-1})]\right\}. \nonumber
\end{align}
This equation is derived in detail in Sec.~\ref{sec ambiguities}. 

Next, we turn to the internal energy and first notice that the Hamiltonian of mean force can be simplified to 
\begin{equation}\label{eq HMF splitting}
 H_X^*(\lambda_t) = \sum_{\bb r_n} H_{SA}^*(\lambda_t,\bb r_n)|\bb r_n\rl\bb r_n| + V_{AI}(\lambda_t).
\end{equation}
This follows from the facts that the IDF are energetically degenerate and that the measurement of the ancilla 
happens after the interaction with the system. That is, at any given time the $k$th ancilla is either in contact with 
the system [and then $V_{A(k)I}(\lambda_t) = 0$] or not, in which case $H_{A(k)} + V_{A(k)I}(\lambda_t)$ commutes with
the rest of the Hamiltonian. Since there is also at most one ancilla in contact with the system at a given time (say 
again the $k$th ancilla), we can also conclude that 
$H_{SA}^*(\lambda_t,\bb r_n) = H_{SA(k)}^*(\lambda_t,\bb r_n) + \sum_{i\neq k} H_{A(i)}$.
In the case of a causal model as considered in Secs.~\ref{sec quantum causal models} 
and~\ref{sec dynamical equivalence} (described by an instantaneous system-ancilla interaction) we can even set 
$H_{SA}^*(\lambda_t,\bb r_n) = H_S^*(\lambda_t,\bb r_n) + \sum_k H_{A(k)}$. The splitting~(\ref{eq HMF splitting}) 
together with Eq.~(\ref{eq state BSAI}) implies for the internal energy [denoting 
$H^*_{SA} = H_{SA}^*(\lambda_t,\bb r_n)$ for simplicity] 
\begin{equation}\label{eq U}
 U(t) = \sum_{\bb r_n} p(\bb r_n) \mbox{tr}\left\{(H_{SA}^* + \partial_\beta H_{SA}^*)\rho_{SA}(t,\bb r_n)\right\}.
\end{equation}
Similarly to the term $V_{XN}(\lambda_t)$ in Eq.~(\ref{eq internal energy pre}), we have also here neglected the 
interaction term $V_{AI}(\lambda_t)$: it describes a very fast process, whose temporary resolution is unimportant 
for us, i.e., for most times $V_{AI}(\lambda_t) = 0$, see Point~II above in Sec.~\ref{sec agreements}. The energetic 
change due to the measurement is nevertheless fully captured by the work~(\ref{eq work AI}). 

Finally, we look at the definition of entropy, Eq.~(\ref{eq S}). Due to Eqs.~(\ref{eq state BSAI}) 
and~(\ref{eq HMF splitting}) this can be written as 
\begin{equation}
 \begin{split}\label{eq S 2}
  S(t) =&~ \sum_{\bb r_n} p(\bb r_n) \{S_\text{vN}[\rho_{SA}(t,\bb r_n)] - \ln p(\bb r_n)\} \\
  &+ \sum_{\bb r_n} p(\bb r_n)\beta^2\mbox{tr}\{(\partial_\beta H_{SA}^*)\rho_{SA}(t,\bb r_n)\}.
 \end{split}
\end{equation}
Similarly, the nonequilibrium free energy~(\ref{eq free energy}) becomes 
\begin{align}
 F(t) =& \sum_{\bb r_n} p(\bb r_n) \mbox{tr}\{H_{SA}^*\rho_{SA}(t,\bb r_n)\} \label{eq F} \\
        & +T\sum_{\bb r_n} p(\bb r_n) \{\ln p(\bb r_n) - S_\text{vN}[\rho_{SA}(t,\bb r_n)]\}. \nonumber
\end{align}

%%%%%%%%%%%%%%%%%%%%%%%%%%%%%%%%%%%%%%%%%%%%%%%%%%%%%%%%%%%%%%%%%%%%%%%%%%%%%%%%%%%%%%%%%%%%%%%%%%%%%%%%%%%%%%%%%%%%%%%%
\subsection{Conditional thermodynamics: the canonical choice}
\label{sec canonical choice}

Let us repeat our philosophy so far: We started with a quantum causal model and constructed 
an autonomous model, which simulates it. The unitary dilation theorem~(\ref{eq Stinespring}) as well as its 
extension~(\ref{eq Stinespring instruments}) to non-deterministic interventions naturally forced us to introduce 
a stream of ancillas and a classical memory into the picture. Then, we studied the thermodynamics of the isolated 
autonomous model by combining recently developed tools in strong coupling 
thermodynamics~\cite{EspositoLindenbergVandenBroeckNJP2010, 
TakaraHasegawaDriebePLA2010, SeifertPRL2016, StrasbergEspositoPRE2019, StrasbergPRL2019} and simplified the resulting 
expression as much as possible. Now, we imagine the same situation as in Sec.~\ref{sec dynamical equivalence} where an 
external super-observer measures the memory and obtains outcome $\bb r_n$. What is the internal energy and system 
entropy as well as the work supplied and the heat flow conditioned on this outcome? 

Above, we already wrote down all thermodynamic quantities in a suggestive way as an ensemble average over 
$\bb r_n$ via 
\begin{equation}
 X(t) = \sum_{\bb r_n} p(\bb r_n) x(\bb r_n,t),
\end{equation}
where $X$ is a placeholder for $W, U, Q$ and $S$.
Therefore, to get the right thermodynamic quantity $X(t)$ on average, $x(\bb r_n,t)$ presents its stochastic 
counterpart (denoted by a small letter as customarily done in stochastic thermodynamics). For instance, the 
stochastic work at the trajectory level follows from Eqs.~(\ref{eq work S}),~(\ref{eq work SA}) and~(\ref{eq work AI}) 
as
\begin{align}
 & w_S(t,\bb r_n) = \label{eq work S stochastic} \\
 &~~~~~\int_{t_0}^t ds \mbox{tr}\left\{\frac{\partial H_S(\lambda_s)}{\partial s}\rho_S(s,\bb r_n)\right\}, \nonumber \\
 & w_{SA}(t,\bb r_n) = \label{eq work ctrl} \\ 
 &~~~~~\int_{t_0}^t ds \mbox{tr}\left\{\frac{\partial V_{SA}(\lambda_s)}{\partial s}\rho_{SA}(s,\bb r_n)\right\}, \nonumber \\
 & w_{AI}(t,\bb r_n) = \label{eq problematic part} \\ 
 &~~~~~\sum_{k=0}^n \mbox{tr}\left\{H_{A(k)}[\rho''_{A(k)}(\bb r_k) - \rho'_{A(k)}(\bb r_{k-1})]\right\}. \nonumber
\end{align}
Likewise, the internal energy~(\ref{eq U}) and heat~(\ref{eq Q}) at the trajectory level become 
\begin{align}
 u(t,\bb r_n) =&~ 
 \mbox{tr}\{(H_{SA}^* + \beta\partial_\beta H_{SA}^*) \rho_{SA}(t,\bb r_n)\}, \\
 q(t,\bb r_n) =&~ u(t,\bb r_n) - w(t,\bb r_n), \label{eq heat stochastic}
\end{align}
where $w(t,\bb r_n) = w_S(t,\bb r_n) + w_{SA}(t,\bb r_n) + w_{AI}(t,\bb r_n)$. 
Finally, the entropy and nonequilibrium free energy follow from Eqs.~(\ref{eq S 2}) and~(\ref{eq F}):
\begin{align}
 s(t,\bb r_n) =& -\ln p(\bb r_n) + S_\text{vN}[\rho_{SA}(t,\bb r_n)] \\
 &+ \beta^2\mbox{tr}\{(\partial_\beta H_{SA}^*)\rho_{SA}(t,\bb r_n), \nonumber \\
 f(t,\bb r_n) =&~ \mbox{tr}\{H_{SA}^* \rho_{SA}(t,\bb r_n)\} \\
 & + T\ln p(\bb r_n) - TS_\text{vN}[\rho_{SA}(t,\bb r_n)]. \nonumber
\end{align}

These quantities, which were derived from an inclusive, Hamiltonian approach, can now be compared with the proposed 
definitions in Refs.~\cite{StrasbergPRE2019, StrasbergPRL2019} (Ref.~\cite{StrasbergWinterPRE2019} deals with the 
classical counterpart). To compare them, one has to keep in mind that the definitions in Ref.~\cite{StrasbergPRE2019} 
were proposed for the weak coupling regime. This implies $H_X^* = H_X$ and in particular $\partial_\beta H_X^* = 0$. 
Furthermore, the ancillas were called `units' in Refs.~\cite{StrasbergPRE2019, StrasbergPRL2019}. The $k$th unit was 
denoted by $U(k)$ and the entire string of units was denoted $U(\bb n)$ instead of $A$. 

Apart from one minor exception, \emph{all definitions coincide}. Therefore, the question raised in 
Ref.~\cite{StrasbergPRE2019} ``whether there exist good \emph{a priori} arguments'' (in contrast to the many 
\emph{a posteriori} justifications given in Refs.~\cite{StrasbergPRE2019, StrasbergWinterPRE2019, StrasbergPRL2019}) 
to justify the definitions used in operational quantum stochastic thermodynamics can be unequivocally be answered 
with ``Yes!'' 

The exception concerns Eq.~(\ref{eq problematic part}), which was previously interpreted as a \emph{heat} exchange of 
the ancilla during the control operation, see, e.g., Eq.~(31) in Ref.~\cite{StrasbergPRE2019} or Eq.~(13) in the 
Supplement of Ref.~\cite{StrasbergPRL2019}. Within our autonomous approach we now recognize it actually as a \emph{work} 
cost, see also below for more details. Interestingly, somewhat anticipating this case, this term was already 
\emph{excluded} from the second law in Refs.~\cite{StrasbergPRE2019, StrasbergPRL2019}. Therefore, no major conclusion 
has to be changed apart from relabeling one term as work instead of heat. In fact, typically this term is of minor 
relevance as it vanishes, for instance, if the ancillas are energetically neutral or if the final measurement of them 
happens in their energy eigenbasis as in Refs.~\cite{SayrinEtAlNature2011, ZhouEtAlPRL2012}. 

%%%%%%%%%%%%%%%%%%%%%%%%%%%%%%%%%%%%%%%%%%%%%%%%%%%%%%%%%%%%%%%%%%%%%%%%%%%%%%%%%%%%%%%%%%%%%%%%%%%%%%%%%%%%%%%%%%%%%%%%
\subsection{Ambiguities in stochastic work and heat}
\label{sec ambiguities}

We first catch up on the promised derivation of Eq.~(\ref{eq work AI}) by focusing on the measurement of the $k$th 
ancilla. During that measurement, described by the interaction Hamiltonian $V_{AI(k)}(\lambda_t)$, the ancilla $A(k)$ 
and the IDF are isolated. The change in their internal energy is therefore identical to the work supplied to them, i.e., 
\begin{widetext}
\begin{equation}\label{eq work AI k start}
 W_{AI(k)} = \sum_{\bb r_{k-1}} p(\bb r_{k-1}) 
 \mbox{tr}\{H_{A(k)}[\C U_\text{meas}^{(k)} - \C I]\rho'_{A(k)}|1,\bb r_{k-1}\rl 1,\bb r_{k-1}|\}.
\end{equation}
Here, we used Eq.~(\ref{eq state BSAI}) and that the IDF are energetically degenerate such that we only have 
to track the change in expectation value of $H_{A(k)}$. Remember that $\rho'_{A(k)} = \rho'_{A(k)}(\bb r_{k-1})$ 
denotes the state of the $k$th ancilla after the interaction with the system, which can depend on $\bb r_{k-1}$. Next, 
we use Eq.~(\ref{eq measurement ancilla}) and take the trace over the IDF to infer that 
\begin{equation}
  W_{AI(k)} = \sum_{\bb r_k} p(\bb r_{k-1}) \mbox{tr}\left\{H_{A(k)} P(r_k)\rho'_{A(k)}P(r_k)\right\}
             - \sum_{\bb r_{k-1}} p(\bb r_{k-1}) \mbox{tr}_{A(k)}\left\{H_{A(k)}\rho'_{A(k)}\right\}.
\end{equation}
Notice that $P(r_k)\rho'_{A(k)}P(r_k) = P(r_k)\rho'_{A(k)}(\bb r_{k-1})P(r_k)$ is a non-normalized state and its 
norm is the probability $p(r_k|\bb r_{k-1})$ to obtain result $r_k$ given the previous results $\bb r_{k-1}$. Thus, by 
writing $P(r_k)\rho'_{A(k)}(\bb r_{k-1})P(r_k) = p(r_k|\bb r_{k-1}) \rho''_{A(k)}(\bb r_k)$,
where $\rho''_{A(k)}(\bb r_k)$ denotes the normalized state of the $k$th ancilla after the measurement conditioned on 
$\bb r_k$, we obtain 
\begin{equation}
 W_{AI(k)} = \sum_{\bb r_k} p(\bb r_k) 
 \mbox{tr}\left\{H_{A(k)} [\rho''_{A(k)}(\bb r_k) - \rho'_{A(k)}(\bb r_{k-1})]\right\}.
\end{equation}
Here, we also used the elementary rules of probability theory $p(\bb r_k) = p(r_k|\bb r_{k-1})p(\bb r_{k-1})$ and 
$\sum_{r_k} p(\bb r_k) = p(\bb r_{k-1})$. This concludes the derivation of Eq.~(\ref{eq work AI}). Consequently, the 
stochastic work~(\ref{eq problematic part}) was identified with the term following $p(\bb r_k)$ in 
Eq.~(\ref{eq work AI}) and the heat~(\ref{eq heat stochastic}) is indirectly defined via the first law. 

We are now in a position, where we can see the origin of the ambiguity in assigning heat and work at the trajectory 
level. Imagine we start with Eq.~(\ref{eq work AI k start}) again, but we express it as
\begin{equation}
 W_{AI(k)} = \sum_{\bb r_{k-1}} p(\bb r_{k-1}) \mbox{tr}\{H_{SA}[\C U_\text{meas}^{(k)} - \C I]\rho'_{SA}|1,\bb r_{k-1}\rl 1,\bb r_{k-1}|\},
\end{equation}
where $H_{SA} = H_S(\lambda_t) + H_A$ is the Hamiltonian of the system and all ancillas. This is possible since 
the operation $\C U_\text{meas}^{(k)}$ acts only non-trivially on the $k$th ancilla and the IDF and hence, the 
expectation value remains unchanged when including additional degrees of freedom. If we then follow the same steps 
as above, we end up with 
\begin{equation}\label{eq work AI k alternative}
 W_{AI(k)} =  
 \sum_{\bb r_k} p(\bb r_k) \mbox{tr}\left\{[H_S(\lambda_t) + H_{A}][\rho''_{SA}(\bb r_k) - \rho'_{SA}(\bb r_{k-1})]\right\}.
\end{equation}
\end{widetext}
Since this expression is still correct, it allows us to confirm by comparison with Eq.~(\ref{eq work AI k start}) 
that the \emph{average} work injected into the system or the remaining ancillas is zero as expected. However, if we now 
follow the strategy $X(t) = \sum_{\bb r_n} p(\bb r_n) x(\bb r_n,t)$ to identify the stochastic work, we obtain the 
definition 
\begin{align}
 & \tilde w_{AI(k)}(\bb r_k) = \label{eq work S stochastic alternative} \\
 & ~~\mbox{tr}\left\{[H_S(\lambda_t) + H_{A}][\rho''_{SA}(\bb r_k) - \rho'_{SA}(\bb r_{k-1})]\right\}. \nonumber
\end{align}
Now, the \emph{stochastic} work injected into the system or the remaining ancillas is not zero since our state of 
knowledge about those entities changes when receiving the measurement result $r_k$. Hence, if we sum this over all 
measurements $k$, we do not get back Eq.~(\ref{eq problematic part}). Consequently, via the first law we also 
get a different expression for the stochastic heat~(\ref{eq heat stochastic}). 

Note that this ambiguity of assigning stochastic heat and work \emph{only} happens during the measurement step of the 
ancilla, i.e., Eqs.~(\ref{eq work S stochastic}) and~(\ref{eq work ctrl}) remain \emph{unchanged}, and it also does 
\emph{not} affect the definitions of state functions such as stochastic internal energy or entropy. 

%%%%%%%%%%%%%%%%%%%%%%%%%%%%%%%%%%%%%%%%%%%%%%%%%%%%%%%%%%%%%%%%%%%%%%%%%%%%%%%%%%%%%%%%%%%%%%%%%%%%%%%%%%%%%%%%%%%%%%%%
\subsection{Comparison with other choices in the literature}
\label{sec thermo other choices}

Together with the section above we are now in a position to reconsider other choices in the literature. In particular, 
the question of how to thermodynamically describe a projective measurement of a quantum system has gained a lot of 
attention. For that particular class of interventions it is actually superfluous to consider the stream of ancillas and 
one could directly look at an interaction between the system and the $k$th IDF to implement a projective measurement as 
described in Point~F of Sec.~\ref{sec autonomous model}. On the other hand, nothing will change in our conclusions 
if we keep the ancilla but simply assume that it is energetically degenerate, i.e., $H_A \sim 1_A$ for the rest of 
this section. 

We start with the two-point projective measurement scheme~\cite{EspositoHarbolaMukamelRMP2009, 
CampisiHaenggiTalknerRMP2011}, which is a theoretically successful approach to derive quantum fluctuation theorems. 
In there, one considers an isolated system subjected to two projective measurements of the energy at the beginning and 
at the end of the protocol. The difference in the measurement outcomes is interpreted as the stochastic work in this 
framework. This stochastic work includes two terms. One term is due to changing the system Hamiltonian $H_S(\lambda_t)$ 
in time, which is fully captured by Eq.~(\ref{eq work S stochastic}). The other term interpretes the change in energy 
caused by updating our state of knowledge due to the final projective measurement as work, which corresponds to the 
alternative choice~(\ref{eq work S stochastic alternative}). Adding these two contributions, results in
\begin{equation}
 \begin{split}
  w_S(t,&\bb r_{n-1}) + \tilde w_{AI(k)}(\bb r_n) \\
  &= \int_{t_0}^t ds \mbox{tr}\left\{\frac{\partial H_S(\lambda_t)}{\partial s}\rho_S(s,\bb r_{n-1})\right\} \\
  &~~~+ \mbox{tr}\{H_S(\lambda_t)[\rho''_S(\bb r_n) - \rho'_S(\bb r_{n-1})]\}.
 \end{split}
\end{equation}
Now, we specialize to the two-point projective measurement scheme, where $\bb r_n = (E_0,E_1)$ only denotes the two 
results of the initial and final projective measurement. The corresponding eigenstates of the initial and final 
Hamiltonian are denoted as $|E_0\rangle$ and $|E_1\rangle$ and we identify $\rho''_S(E_0,E_1) = |E_1\rl E_1|$ and 
$\rho'_{S}(E_0) = U_S(t_1,t_0)|E_0\rl E_0|U_S^\dagger(t_1,t_0)$ denotes the unitarily evolved system state prior to the 
final measurement. Since the system is assumed to be isolated here, it follows that 
\begin{equation}\label{eq work S TPM}
 w_S(t,\bb r_{n-1}) + \tilde w_{AI(k)}(\bb r_n) = E_1 - E_0. 
\end{equation}
Therefore, Eq.~(\ref{eq work S TPM}) reproduces the work statistics of the two-point projective measurement approach.

Does this imply that Eq.~(\ref{eq work S stochastic alternative}) is the natural choice instead of 
Eq.~(\ref{eq problematic part})? Notice that in this paper we were mainly interested in an \emph{open} system coupled 
to a heat bath. Now, suppose we were to follow the ideology of the two-point projective measurement approach and 
consider the following example. At some initial time $t_0$ we have prepared a two-level system with energy gap $\Omega$ 
in its excited state, $\rho_S(t_0) = |e\rl e|$, which then evolves in time while being in contact with a heat bath 
(which, for the sake of simplicity, is considered to be an ideal weakly coupled Markovian heat bath here). Then, we 
perform at time $t_1>t_0$ a measurement of its energy and find it in the ground state $|g\rangle$. If we do not drive 
the system ($\lambda_t =$ constant), its change in internal energy is simply 
\begin{equation}
 \Delta u = \lr{g|H_S|g} - \lr{e|H_S|e} = -\Omega.
\end{equation}
Clearly, a natural interpretation of this situation would suggest to identify $\Delta u$ with the \emph{heat} exchanged 
with the bath, which induced at some unknown time $t\in(t_0,t_1)$ a jump from the excited to the ground state. Instead, 
the two-point projective measurement approach would identify parts of $\Delta u$ as work, namely the part of energy 
change caused by a change of its state from $\rho'_S(t_1)$ (the state prior to the measurement at $t_1$) to $|g\rl g|$ 
(the post-measurement state), cf.~Eq.~(\ref{eq work S stochastic alternative}). For open quantum systems, the two-point 
projective measurement approach therefore does not reproduce our classical intuition about heat exchanges induced by 
stochastic transitions from one state to another, which are revealed by updating our state of knowledge. In fact, one 
can show that the canonical choice of Sec.~\ref{sec canonical choice} reduces to the conventional definitions used in 
classical stochastic thermodynamics~\cite{SekimotoBook2010, SeifertRPP2012, VandenBroeckEspositoPhysA2015} when 
considering ideal continous measurements of an open classical system~\cite{StrasbergPRE2019}. 
Furthermore, note that Eq.~(\ref{eq work S TPM}) excludes the energetic cost of the first measurement yielding 
result $E_0$. However, if the system was prior to the measurement in weak contact with a heat bath and only 
afterwards isolated, Eq.~(\ref{eq work S TPM}) is the correct stochastic work if one adapts the convention that 
Eq.~(\ref{eq problematic part}) is the correct choice for open quantum systems.

An opposite interpretation to the two-point projective measurement approach was suggested in 
Ref.~\cite{ElouardEtAlQInf2017}, where the change in energy of an isolated system due to a projective measurement of an 
arbitrary observable was identified as heat. This heat does not appear in any second law and it was called ``quantum 
heat''. While we see that our canonical choice in Sec.~\ref{sec canonical choice} allows to identify parts of the 
changes in energy due to a projective measurement as heat, \emph{on average} it predicts that any change in energy due 
to a measurement is due to work, which follows from Eq.~(\ref{eq work AI k alternative}). This average, derived within 
our inclusive, Hamiltonian approach, agrees with the two-point projective measurement approach on average and coincides 
with the ``switching work'' known from the repeated interaction framework~\cite{BarraSciRep2015}, see also 
Ref.~\cite{StrasbergEtAlPRX2017}. Therefore, the concept of ``quantum heat'', at least as originally introduced in 
Ref.~\cite{ElouardEtAlQInf2017}, does not have any theoretical foundation within our autonomous approach. 

%%%%%%%%%%%%%%%%%%%%%%%%%%%%%%%%%%%%%%%%%%%%%%%%%%%%%%%%%%%%%%%%%%%%%%%%%%%%%%%%%%%%%%%%%%%%%%%%%%%%%%%%%%%%%%%%%%%%%%%%
\section{Final remarks}
\label{sec remarks}

The main message of this paper is a very positive one. After 20 years of debate, the present paper shows that there 
exists a straightforward way to derive the definitions of quantum stochastic thermodynamics by starting from unambiguous 
notions at the unmeasured level. A certain amount of freedom in defining heat and work at the stochastic level 
remains, but additional arguments can be invoked in favour of one or the other. In particular, the most consistent 
choice might depend on the question whether the considered system is open or isolated. That this can give rise to 
different thermodynamic definitions should not be too surprising as this is the same in classical 
thermodynamics. Furthermore, the resulting definitions turn out to be surprisingly simple and mostly follow from what 
was known (since a long time) at the unmeasured level \emph{if} one correctly takes into account the measurement results 
$\bb r_n$. This basically means that one has to replace $\rho_S(t)$ by the correct state of knowledge 
$\rho_S(t,\bb r_n)$ to compute, e.g., the stochastic work or internal energy. 

Furthermore, it cannot be overemphasized that the operational framework of quantum stochastic thermodynamics equips 
a large class of quantum causal models with a consistent thermodynamic interpretation, even 
along a single trajectory. The main physical assumptions is an initially equilibrated system-bath state, 
the remaining assumptions listed at the end of Sec.~\ref{sec quantum causal models} are rather of minor relevance 
for current practical purposes in quantum thermodynamics. 
In particular, the present paper shows that the strong coupling definitions even hold in case of real-time feedback 
control, which could not be established in Ref.~\cite{StrasbergPRL2019}. Thus, operational quantum stochastic 
thermodynamics opens up the possibility to analyse the thermodynamics of almost every quantum experiment, even beyond 
average quantities, and its thermodynamic consistency is guaranteed by virtue of the results reported here. 

There is one caveat, however, which is not linked to the framework of operational quantum stochastic thermodynamics 
\emph{per se} but rather to the limit in which a quantum causal model or quantum stochastic process is defined. 
As long as the system-ancilla interaction is not instantaneous, a clear advantage of operational quantum stochastic 
thermodynamics is that it allows to define thermodynamic quantities, even along a single trajectory, solely in terms 
of experimentally available information. Everything can be computed based on knowledge of the conditional 
system-ancilla state $\rho_{SA}(t,\bb r_n)$ given a trajectory of measurement results $\bb r_n$. In this sense, the 
theory is fully `operational'. But, quite ironically, this is no longer true in the peculiar limit, where the 
system-ancilla interaction $V_{SA(k)}(\lambda_t)$ is idealized as a delta-peak 
[see Eq.~(\ref{eq instantaneous interaction})]. This implements a unitary 
$\C U_\text{ctrl}^{(k)}$ on the system-ancilla space, whose energetic change is work. But if the 
system-bath coupling $V_{SB}$ is not negligible, the work $W_{SA}(t_k)$ invested in the $k$th control operation becomes 
\begin{align}
 &W_{SA}(t_k) \label{eq work ctrl singular} \\
 &~~~= \mbox{tr}\left\{H_\text{tot}(\lambda_k)(\C U_\text{ctrl}^{(k)} - \C I)\rho_\text{tot}(t_k)\right\} \nonumber \\
 &~~~= \mbox{tr}\left\{[H_S(\lambda_k)+V_{SB}+H_{A(k)}](\C U_\text{ctrl}^{(k)} - \C I)\rho_{SBA}(t_k)\right\}. \nonumber
\end{align}
This shows that one has to eventually evaluate the term 
$\mbox{tr}_{SBA}\{V_{SB}(\C U_\text{ctrl}^{(k)} - \C I)\rho_{SBA}(t_k)\}$, which requires explicit knowledge about the 
bath degrees of freedom, albeit for any smooth, non-singular time-dependence of $V_{SA(k)}(\lambda_t)$ this is never 
necessary, see Eq.~(\ref{eq work ctrl}). Thus, beyond the weak coupling regime, the strict limit of a quantum causal 
model makes the operational approach no longer fully operational. However, at least for typical open quantum systems 
linearly coupled to a quadratic bath, Eq.~(\ref{eq work ctrl singular}) can be still efficiently computed using reaction 
coordinate master equations as explicitly demonstrated in, e.g., Refs.~\cite{StrasbergEtAlNJP2016, 
NewmanMintertNazirPRE2017, StrasbergEtAlPRB2018}. 

\emph{Acknowlegdements.} I am grateful to Jens Eisert for raising the question whether the operational framework of 
quantum stochastic thermodynamics can be derived from an autonomous perspective. This research was financially 
supported by the DFG (project STR 1505/2-1) and the Spanish MINECO FIS2016-80681-P (AEI-FEDER, UE). 

%%%%%%%%%%%%%%%%%%%%%%%%%%%%%%%%%%%%%%%%%%%%%%%%%%%%%%%%%%%%%%%%%%%%%%%%%%%%%%%%%%%%%%%%%%%%%%%%%%%%%%%%%%%%%%%%%%%%%%%%

\bibliographystyle{apsrev4-1}
%\bibliography{/home/philipp/Documents/references/books,/home/philipp/Documents/references/open_systems,/home/philipp/Documents/references/thermo,/home/philipp/Documents/references/info_thermo,/home/philipp/Documents/references/general_QM,/home/philipp/Documents/references/math_phys,/home/philipp/Documents/references/equilibration}
\bibliography{/home/wiwi/Documents/references/books,/home/wiwi/Documents/references/open_systems,/home/wiwi/Documents/references/thermo,/home/wiwi/Documents/references/info_thermo,/home/wiwi/Documents/references/general_QM,/home/wiwi/Documents/references/math_phys,/home/wiwi/Documents/references/general_refs,/home/philipp/Documents/references/equilibration}

\end{document}